\documentclass[aps,prl,twocolumn,groupedaddress,floatfix,showpacs]{revtex4}

\usepackage{subfigure}
\usepackage{color}
\usepackage{graphicx}
\begin{document}

\title{Chemically active substitutional nitrogen impurity in carbon nanotubes}



\author{Andriy H. Nevidomskyy, G\'abor Cs\'anyi, and Michael C. Payne}


\affiliation{Theory of Condensed Matter group, Cavendish Laboratory, University
of Cambridge, Cambridge CB3 0HE, UK}


\date{\today}

\begin{abstract}
We investigate the nitrogen substitutional impurity in semiconducting
zigzag and metallic armchair single--wall carbon nanotubes using
\emph{ab initio} density functional theory.  At low concentrations
(less than 1 at. \%), the defect state in a semiconducting tube
becomes spatially localized and develops a flat energy level in the
band gap.  Such a localized state makes the impurity site chemically
and electronically active.  We find that if two neighbouring tubes
have their impurities facing one another, an inter--tube covalent bond
forms.  This finding opens an intriguing possibility for tunnel
junctions, as well as the functionalization of suitably doped carbon
nanotubes by selectively forming chemical bonds with ligands at the
impurity site. If the intertube bond density is high enough, highly
packed bundle of interlinked single--wall nanotubes can form.
\end{abstract}

%\keywords{}

\maketitle

Ever since their discovery\cite{1976nanotube,iijima}, carbon nanotubes
(CNT) have been heralded as the new wonder--material of the future.
Their remarkable mechanical and electronic properties destine them
to play a major role in all kinds of nanotechnologies and molecular
electronics.  At least two major hurdles have to be overcome in order
to fulfill this potential.  First, manipulation of individual tubes
is at best difficult today, which prevents mass production of devices.
Second, the ability to fine--tune the various properties of the material
to suit particular applications has to be achieved.

Carbon nanotubes exhibit a variety of electronic properties, for
example, depending on their diameter and chirality, they vary from
being metallic to semiconducting\cite{Dresselhaus}. The presence of defects and
impurities that are electronically or chemically active can change
these properties and thus have a significant bearing on a broad range
of applications. In this paper we investigate the nitrogen
substitutional impurity in single--wall carbon nanotubes (SWNT).  In the case
of semiconducting tubes, doping with electrons or holes is the
principle route to making electronic devices.  Alternatively,
introducing new levels in the band gap with associated electronic
states that are spatially localized, can create chemically active
impurity sites.

Several groups have produced nanotubes containing nitrogen
\cite{Yudasaka1997,Sen98,Terrones1999,Grobert2000,Han2000,Kurt2001,
Czerw2001,Terrones2002,Trasobares2002,Wang2002,Lee2002} or nitrogen
and boron\cite{Sen98,Stephan1994,GolbergSWBCN}.  All of these were
multi wall tubes, except for \cite{GolbergSWBCN}, where nitrogen was
always accompanied by boron.  The measured atomic concentrations of
the impurities were relatively high in all cases, above 1\%, and all
the CN$_{x}$ tubes were found to be metallic, as predicted by theoretical
models\cite{Terrones2002,Kaun2002,Czerw2001}.  Indeed, one other
reason for the particular interest in CN$_{x}$ tubes was that they
might make even initially semiconducting tubes metallic.

Various structural models have been proposed for the incorporation of
nitrogen into the carbon network.  Direct substitution was first
studied by Yi and Bernholc\cite{Yi} using density functional theory at
an atomic impurity concentration of about 1\%, where they found an
impurity state lying 0.27 eV below the bottom of the conduction band
at the $\Gamma$ point and a significant overlap of the states
associated with the adjacent nitrogens (8.4 \AA\ apart).  Lammert
\emph{et al.}\cite{Lammert2001} considered disordered substitution of
boron and nitrogen at a concentration of 3.5\% and 1.4\%, and observed
both donor and acceptor levels in the gap.  For an isolated nitrogen
or boron impurity, they found the impurity state to be localized
within about 10~\AA\ for an (8,0) tube.

Another atomic arrangement for the experimentally observed nitrogen
impurity was proposed by Terrones \emph{et al.}\cite{Terrones1999},
where they suggest that at high concentrations (above about 10\%)
nitrogens are divalent, pyridine--like.  Later publications from the
same group refine this model and correlate the calculated density of
states of metallic tubes with experimental observations
\cite{Czerw2001,Terrones2002}.

To study the isolated substitutional nitrogen defect, the simulation
cell has to be as long as possible. The existing constraints of
computational power mean that the diameter of the tube should
therefore be as small as possible. In all calculations we used the
achiral zigzag (8,0) and armchair (5,5) nanotubes.  Our simulation
cell has periodic boundary conditions, the tubes are arranged parallel
to the z axis and form a triangular lattice in the perpendicular x--y
plane.  This arrangement is experimentally seen in carbon nanotube
ropes\cite{Dresselhaus}.  We studied a range of nitrogen
concentrations from 0.26\% to 1\% by introducting one impurity atom
and changing the size of the simulation cell along the axis of the
tube. The largest simulation cell contained 12 unit cells along the z
axis for the zigzag tube and 14 unit cells for the armchair tube,
respectively, with one impurity atom in it.

Throughout, we used \emph{ab-initio} plane--wave pseudopotential
density functional theory\cite{payne} as implemented in the CASTEP
code\cite{castep}.  The generalized gradient approximation was used to
account for exchange and correlation in the Perdew--Burke--Ernzerhof
form\cite{gga-PBE}.  All calculations were spin-polarized and ultrasoft
pseudopotentials were used for carbon and nitrogen with a cutoff
energy of 300 eV.  We assessed the adequacy of this cutoff by
performing calculations at 350 eV for the smallest unit cells and
verified that none of properties reported in this work changed
perceptibly.  For each concentration, the ionic positions were
optimized by using one k--point at the $\Gamma $ point for the largest
simulation cells, and correspondingly more k--points for the smaller
cells.  After the geometry optimization has converged, the band
structure was calculated using a finer mesh of k--points.

\begin{figure}[!b]

\begin{center}\includegraphics[  width=8.6cm]{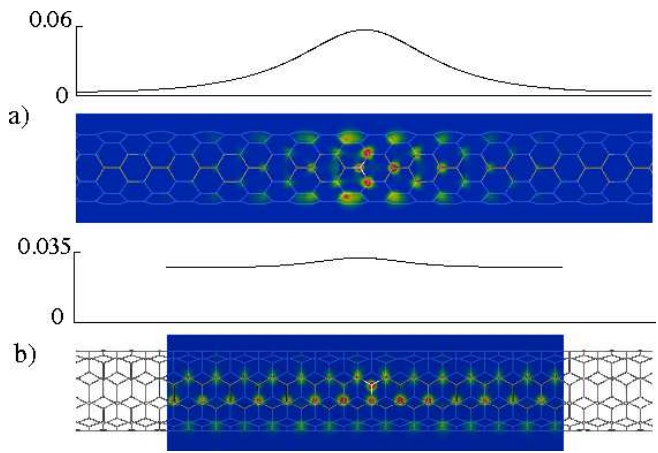}
\end{center}

\caption{\label{spindensity}Slice of the spin density of the N-doped
carbon nanotube in a plane parallel to the tube axis, the plane being
tangential to the tube at the position of the nitrogen ion, for a)
(8,0) zigzag tube, comprising 12 unit cells (50.6 \AA), and b) (5,5)
armchair tube, comprising 14 unit cells (34.3 \AA).  Above each slice,
we plot the envelope of the spin density (integrated in the plane
perpendicular to the tube) in the units of electron/\AA. 
The nitrogen ion is marked white. The
colour range maps the value of the spin density, with blue
corresponding to the smallest and red to the highest values.  Note
that the scale of the spin density is quite different for the two
panels.  }
\end{figure}

We find that the equilibrium position of the nitrogen ion is almost
unchanged with respect to the corresponding C atom in the undoped
nanotube, being moved by at most 0.01 \AA, similarly to the case of
azafullerene, C$_{59}$N\cite{C59N,gabor_C59N}.  A slice of the
calculated spin--density for both zigzag and armchair cases is shown
in Fig. \ref{spindensity}.  The spin density is maximal on the N ion
and the neighbouring C ions and shows oscillatory behaviour as
it reaches a local maximum on every other ion in the network, again
similarly to that observed in C$_{59}$N\cite{C59N,gabor_C59N}.  The
shape of the spin density isosurfaces is reminiscent of the $\pi
$--orbitals in graphite with their axes of symmetry perpendicular to
the sheet. 

The localization of the extra electron associated with the isolated
impurity shows a striking difference between the metallic and
semiconducting cases, despite the fact that the band gap of the latter
is quite small (0.6 eV according to the DFT calculations). In the
armchair tube, the unpaired electron is almost completely delocalized
along the tube, as can be seen from the plot of the envelope of the
spin density in the lower panel of Fig.~\ref{spindensity}.  However,
for the zigzag tube, it is localized near the impurity site, but the
localization region is fairly large, about 30 \AA. This region
contains about 90\% of the unpaired spin density. The decaying
envelope of the spin density is exponential.

The oscillations in the spin density can be understood as follows.
Regarding the nitrogen atom as perturbation, the response of the
system is characterized by the static susceptibility $\chi(r)$.  The
characteristic period in $\chi(r)$ corresponds to a maximum in the
Fourier transform, $\chi(q)$, which happens for $q$ values that
connect points in the Brillouin zone with the minimum energy
difference.  For the 2D graphene lattice, the energy gap is zero at
the high symmetry K points, so for this, and similar systems,
$\chi(r)$ will oscillate with a period corresponding to the momentum
transfer between different K points, which is what is observed.  The
well known oscillations in the susceptibility in the RKKY
formalism\cite{RKKY} are essentially a special case of the above
considerations.

The decay of the envelope of the spin density is also analogous to the
RKKY case. For metals in 1D, the result is $\chi(r) \propto
1/r$ \cite{aristov97, litvinov98}. For semiconductors, an analogous
coupling to excited states takes place, and assuming a parabolic
conduction band, it was shown \cite{rowland55} that $\chi(r) \propto \exp[-(\alpha
E_{\mathrm{gap}}^{1/2}) r]$, where $\alpha = (2
m_{\mathrm{eff}})^{1/2}/\hbar$. For the zigzag
nanotube, using the calculated value of $m_{\mathrm{eff}}/m_e = 0.2$
this crude estimate gives a decay constant of about 6 \AA, close to
the observed 7 \AA.

Note that in all the graphene-based systems (C$_{59}$N, nitrogen doped
graphene and nanotubes), the spin density is positive almost
everywhere.  At first sight, this might seem to contradict the results
from electron spin resonance hyperfine measurements, which indicate
that the spin density oscillates symmetrically around zero.  In fact,
as was mentioned above, the spin density is very strongly dominated by
the $p$ component, whereas the Fermi contact interaction in the
hyperfine coupling is only sensitive to the $s$ component, so its
observed symmetrical oscillation is consistent with the calculations.

A crucial difference between the armchair and the zigzag tubes is that
the introduction of the nitrogen atom breaks the left--right mirror
symmetry in the latter case.  As two impurity atoms approach each
other in an armchair tube, their interaction will show oscillatory
behaviour depending on whether the spin density maxima induced by the
respective nitrogens coincide or not. But in the zigzag tube the local
maxima of the spin density occur on different sublattices on the two
sides of the nitrogen atom.  As two impurities approach, their
respective spin density maxima never coincide, irrespective of whether
the nitrogens occupy the same sublattice or not.

\begin{figure}[!btp]

\begin{center}{\includegraphics[  width=8.6cm]{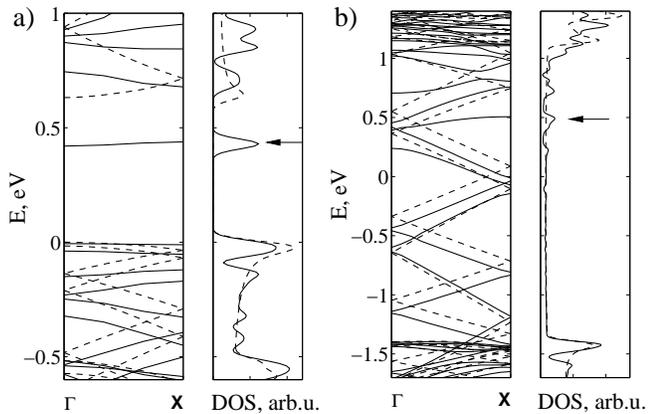}}\end{center}

\caption{\label{bandstructures}Band structures and corresponding densities
of states for the largest N-doped carbon nanotubes studied (solid
line) versus the folded band structure of pure nanotube (dashed line)
in a) (8,0) zigzag nanotube, bands aligned in such a way that the valence
band edge coincides with that of pure nanotube, and b) (5,5) armchair
tube, bands aligned in such a way that the van Hove singularities below the
Fermi level coincide.  The arrows mark the new band associated with
the impurity.  For reasons of clarity we only show the up--spin channel,
which contains the unpaired electron. Note that below the Fermi level,
the splitting of the spin channels is less than 0.03 eV. }
\end{figure}

Figure \ref{bandstructures} shows the band structure of the
nitrogen-doped nanotubes along with the corresponding densities of
states. The band structure of undoped tube (folded in such a way, that
it fits into the shrinked Brillouin zone corresponding to the larger
supercell) is shown for comparison. For the zigzag case, the extra
band of the impurity falls somewhere near the bottom of the conduction
band, and a complicated hybridization takes place with the existing
unoccupied bands.  The end result is that a new flat level appears
about 0.2 eV below the bottom of the conduction band.  In the armchair
tube, the impurity introduces a peak at about 0.6 eV below the first
van Hove singularity above the Fermi level, as well as a number of
additional peaks, which are formed because of the hybridization and
level repulsion taking place between the energy levels of impurity and
pristine nanotube.

\begin{figure}[!btp]

\includegraphics[  width=8.6cm]{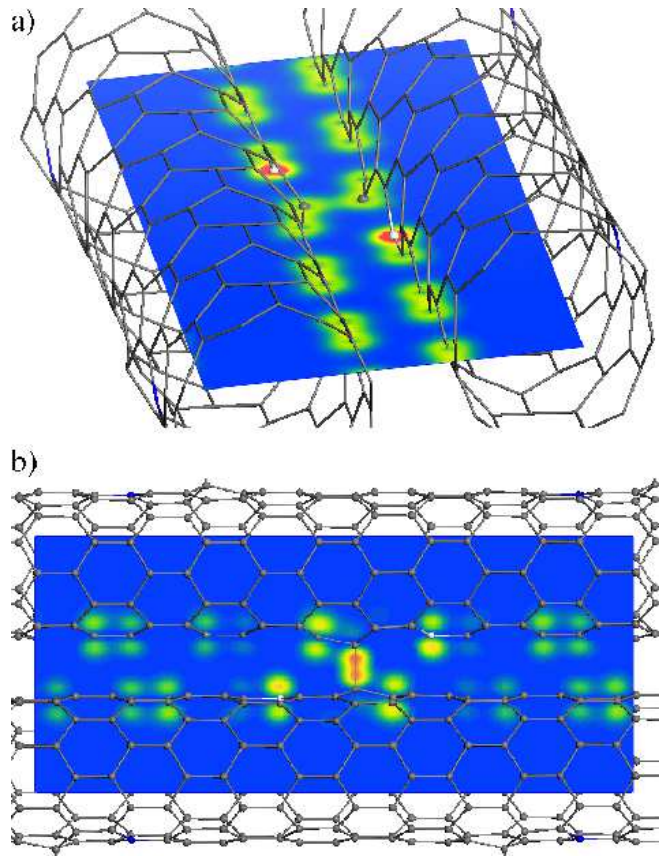}

\caption{\label{tube-tube charge}Two zigzag nanotubes each doped with
nitrogen (white balls), facing each other. The chemical bond is formed
between the two carbon atoms (grey balls), which have the maximum of
the spin density in the isolated nanotube. The slice taken through the
C--C bond shows the distribution of the a) total charge density and b)
density of the HOMO orbital, using the colour map, with blue
corresponding to the lowest and red to the highest densities.}

\end{figure}

The localized electronic state of the semiconducting tube offers the
intriguing possibility of being chemically active and forming covalent
bonds with ligands or other tubes. Figure \ref{tube-tube charge} shows
two zigzag nanotubes each doped with nitrogen (one per four unit
cells) in such a way that two carbon atoms with originally significant
spin density on them face each other.  As the charge density slice
shows, a covalent bond is formed between these atoms, and thus between
the tubes.  The two carbon atoms pop out of the tangent plane of the
tubes, and take up an $sp^{3}$ configuration, the bond length between
them being 1.65 \AA (note that a similar geometry was discussed in the
context of carbon nitride materials\cite{sandre}). In order to obtain
the geometry shown, the two tubes were pressed against each other,
which is realistic when modelling a nanotube lying on top of another
on a surface\cite{tubetubetunneling}.  The further corroboration of
the formation of a true covalent bond between the carbon atoms comes
from the analysis of the charge density originating from different
orbitals. The largest contribution to the total charge density between
the newly bonded carbon atoms comes from the HOMO orbital (Fig.
\ref{tube-tube charge}b), which is occupied by about 0.3 of an
electron.

On the other hand, for the intertube bonds to be realized in a bundle
of SWNTs, they have to be stable under zero external pressure.  We
found that this is possible if the number of intertube bonds is
increased to one per two unit cells (8.4 \AA), so that enough energy
is gained in the new bonds to overcome the intertube repulsion. Here,
the bonded configuration is a local minimum and the tube-tube distance
is reduced to 2.5 \AA, from 3.4 \AA in the case of the unbonded
bundle.  This finding opens the possibility of producing highly packed
crosslinked bundles of doped nanotubes; we obtained a crude estimate
of 80 GPa for the bundle shear modulus (as a result of just one
interlink for every 2 unit cells), which is substantially larger than
1-6 GPa, reported for ordinary bundles\cite{Forro}.

%%%%%%%%%%%%%%%%%%%%%%%%%%%%%%%%%
\begin{figure}[!btp]
\begin{center}{\includegraphics[width=8.6cm]{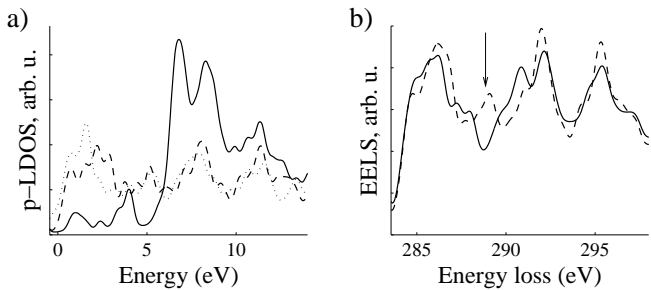}} 
                                            %{LDOS_EELS.eps}} 
                                            %{combine_LDOS_EELS_new.eps}}
\end{center}

\caption{\label{LDOS} a) $p$ projections of local densities of states
 at different atoms: $sp^3$ carbon atom in the C--C bond (solid line),
 $sp^2$ carbon atom in the same tube, but far away from the bond
 (dotted line), $sp^2$ carbon atom in the unbonded isolated nanotube
 (dashed line); b) calculated EELS spectrum near the carbon K edge of
 bonded (solid line) and unbonded (dotted line) tubes,
 which is proportional to the $p$ projected total density of states,
 shifted appropriately. }
\end{figure}
%%%%%%%%%%%%%%%%%%%%%%%%%%%%%%%%%%%%

Panel a) of Fig. \ref{LDOS} shows the $p$ projected local density of
states (DOS) of various carbon atoms in different sections of the
nanotube with one intertube bond per 2 unit cells, as compared to that
of the unbonded system.  The dashed curve shows the fine structure of
the $\pi^*$ peak, which is almost completely absent for the $sp^3$
carbon atom (solid curve) that takes part in the intertube bond.  As a
first approximation, the $p$ projected total DOS in the conduction
band, when suitably shifted, is proportional to the electron energy
loss spectrum (EELS)\cite{egerton_book,rez99}, and shown on panel
b). In the bonded system the $\pi^*$ peak is somewhat diminished, but
more importantly, the peak marked with an arrow is absent.  Since the
proportion of $sp^3$ bonded carbon atoms is expected to be quite
small, a local probe such as EELS would be most suitable for the
detection of the intertube bond.

We performed \emph{ab initio} calculations of zigzag and armchair
carbon nanotubes substitutionally doped with nitrogen.  In the case of
the armchair nanotube, the impurity state is totally delocalized and
the corresponding energy level falls inside the unoccupied bands of
the pristine nanotube.  In the case of the semiconducting zigzag
nanotube, however, the isolated nitrogen impurity forms a flat energy
level lying inside the band gap, 0.2 eV below the bottom of conduction
band. It hybridizes with the $\pi$ orbitals to create a spatially
localized state decaying exponentially with an overall extent of about
30 \AA.  This singly occupied state is chemically active, and we
demonstrate the possibility of the formation of a covalent bond
between two nanotubes with impurities that face one another.  This
opens the intriguing general possibility for chemical bonding between
suitably doped carbon nanotubes or indeed with other ligands. Such
tube--tube junctions could be interesting from a device perspective as
the intertube bond will probably change the tunneling properties
between the tubes considerably.  If the density of intertube bonds is
high enough, a bundle of SWNTs can become highly interlinked and
packed, substantially enhancing its mechanical properties.  The
ligand--binding properties of the localized impurity site, of interest
in chemical sensing applications, are currently under investigation.

\begin{acknowledgments}
The authors would like to thank Peter Littlewood, Ben Simons, David
Khmelnitskii, Malcolm Heggie, Bill Allison and Howard Hughes for
valuable discussions.  Computational work was carried out at the HPCF,
University of Cambridge and supported by grant EC HPRN-CT-2000-00154.
\end{acknowledgments}

\end{document}